\documentstyle[11pt,newpasp,twoside,epsf]{article}
\def\arg#1{{\it#1\/}}

\def\edcomment#1{\iffalse\marginpar{\raggedright\sl#1\/}\else\relax\fi}
\reversemarginpar
\newcommand {\ovi}{O~{\sc vi}}
\newcommand {\nv}{N~{\sc v}}
\newcommand {\civ}{C~{\sc iv}}

\def\eps@scaling{1}
\def\epsscale#1{\gdef\eps@scaling{#1}}

\def\plotone#1{\centering \leavevmode
    \epsfxsize=\eps@scaling\columnwidth \epsfbox{#1}}

\begin{document}
\title{X-ray and UV Views of Hot Gas in Planetary Nebulae}
 \author{You-Hua Chu, Mart\'{\i}n A.\ Guerrero, and Robert A.\ Gruendl}
\affil{Astronomy Department, University of Illinois, 1002 W. Green Street,
Urbana, IL 61801, USA}

\begin{abstract}
The interior of a planetary nebula (PN) is expected to be filled with 
shocked fast wind from the central star.  This hot gas plays the most 
important role in the dynamical evolution of the PN; however, its 
physical conditions are not well-known because useful X-ray and far-UV 
observations were not available until the advent of {\it Chandra}, 
{\it XMM-Newton}, and {\it FUSE}.  This paper reviews X-ray 
observations of the hot gas in PN interiors and far-UV observations of 
the interfaces between the hot gas and the dense nebular shells.

\end{abstract}

\section{Introduction - Origin of Hot Gas in PNe}

It is commonly accepted that PNe are formed by the current fast stellar
wind sweeping up the former slow asymptotic giant branch (AGB) wind.  
In this interacting-stellar-winds scenario (Kwok, Purton, \& 
FitzGerald 1978), the physical 
structure of a PN is similar to that of an interstellar wind-blown
bubble as modeled by Weaver et al.\ (1977).  The interior of a PN is
filled with shocked fast wind reaching temperatures of $10^7-10^8$ 
K, and this hot gas is confined within a 10$^4$ K nebular shell 
consisting of swept-up AGB wind.  The shocked fast wind per 
se has too low a density to emit appreciably in X-rays.  At the 
interface between the hot interior and the cool nebular shell, 
heat conduction lowers the temperature and mass evaporation raises 
the density of the hot interior gas, producing optimal conditions 
for X-ray emission.  Therefore, soft X-ray emission is expected 
from the 10$^6$ K gas near the interface.

The physical processes taking place at the interface have been
mostly a theoretical topic.  The interface physics is expected to
be complicated by the presence of magnetic field and turbulences.
To thoroughly understand the interface physics, observations of 
the transition region are needed.  While the 10$^6$ K gas can be 
observed in X-rays, the adjacent layer of 10$^5$ K gas can 
be observed only in the UV and far-UV through spectral lines of 
collisionally ionized C, N, and O at high ionization stages.

\section{X-ray Views of Hot Gas in Planetary Nebulae}

X-ray observations of the hot gas in PN interiors can be 
confused by the X-ray emission from the central star of a PN 
(CSPN).  A hot CSPN with a low opacity in its atmosphere can be
a soft X-ray emitter, while a CSPN with a late-type companion
may show a coronal X-ray source.  The photospheric emission
from a hot CSPN peaks below 0.1 keV and can be 
spectroscopically discerned from thin plasma emission.
On the other hand, the coronal emission from a late-type 
companion may show similar
spectral shapes to plasma emission; thus, emission from the hot 
gas in PN interiors can be unambiguously confirmed {\em only} if 
it is spatially resolved.  The quest for diffuse X-ray emission 
from hot gas in PN interiors is long and  difficult, but 
significant progress has been made as the sensitivity and 
resolution of X-ray detectors have improved.

\subsection{X-ray Observing Facilities}

Since 1978, several X-ray satellites have been launched 
with imaging instruments sensitive to soft X-rays in the
energy range 0.1--2 keV, where X-ray emission from PNe
is expected.  Table 1 summarizes the X-ray observing 
facilities available in the past 25 years.  The earlier
imaging detectors with spectral capabilities are mostly 
proportional counters with inherently low angular resolution
($\ge30''$) and spectral resolution (e.g., $E/\Delta E \sim 
$ 2.2 at 1 keV for {\it ROSAT} PSPC).  This situation is 
vastly improved by the advent of solid-state and CCD detectors,
e.g., {\it Chandra} ACIS has an angular resolution of 1$''$ 
and a spectral resolution of $E/\Delta E \sim $10 at 1 keV.  
The currently available {\it Chandra} ACIS and {\it XMM-Newton} 
EPIC, with their unprecedented spatial and spectral resolutions 
and sensitivity, are ideal for X-ray observations of PNe.
\begin{table}
\caption{X-ray Observing Facilities Available in Recent Years}
\vspace*{0.1cm}
\begin{tabular}{ccccc}
\tableline\tableline
Satellite & Period of    & Instrument & Angular     & Spectral  \\
Name      & Operation   &            & Resolution  & Resolution \\
\tableline
{\it Einstein} & 1978 -- 1981  & IPC & 120$''$ &  Low \\
               &               & HRI & 5$''$   &  None  \\
{\it EXOSAT}   & 1983 -- 1986  & LE  & 30$''$  &  None  \\
{\it ROSAT}    & 1990 -- 1998  & PSPC & 30$''$ &  Low \\
               &               & HRI & 5$''$   &  None  \\
{\it ASCA}     & 1993 -- 2000  & SIS & 150$''$ &  High \\
{\it Chandra}  & 1999 -- pres. & ACIS & 1$''$  &  High \\
               &               & HRC  & 0\farcs5 & None  \\
{\it XMM-Newton} & 1999 -- pres. & EPIC & 6$''$ & High \\
\tableline\tableline
\end{tabular}
\end{table}

\subsection{X-ray Observations of PNe}

The earliest X-ray observations of PNe were aimed to detect the 
photospheric emission from the CSPNs.  While {\it Einstein}
observed PNe several years before $EXOSAT$, the first report of 
X-ray emission from a CSPN was made with 
{\it EXOSAT} observations of NGC\,1360 by de Korte et al.\ (1985).
Using {\it Einstein} archival data of PNe, Tarafdar \& Apparao 
(1988) reported X-ray emission from NGC\,246, NGC\,6853, NGC\,7293, 
and A\,33.  Similarly, Apparao \& Tarafdar (1989) used the $EXOSAT$ 
archive and reported X-ray emission from NGC\,1535, NGC\,3587, 
NGC\,4361, and A\,36.  The reported emission from A\,33 and 
NGC\,1535 was later shown by {\it ROSAT} observation to arise 
from background X-ray sources (Conway \& Chu 1987; Chu, Gruendl, 
\& Conway 1998).

The search for diffuse X-ray emission from hot gas in PN interiors
was not attempted until {\it ROSAT} became available.  During the 
first six months of the {\it ROSAT} mission, a PSPC detector was 
used to scan the whole sky.  Using this {\it ROSAT} All-Sky Survey,
Kreysing et al.\ (1992) reported diffuse X-ray emission from a number
of PNe.  Unfortunately, their work was plagued by low S/N ratios in 
the data, electronic ghost image of the PSPC detectors at energies 
below 0.2 keV, and mis-identification of nearby or companion coronal 
sources.  After careful scrutiny, the only object in their sample
to show X-ray emission marginally more extended than the point spread 
function of the PSPC was NGC\,6543.

During the {\it ROSAT} mission, pointed observations of a large 
number of PNe were made, but only the few positive detections were
reported, for example,  NGC\,246, NGC\,1360 (Hoare et al.\ 1995), 
NGC\,3587 (Chu, Gruendl, \& Conway 1998), and NGC\,7293 (Leahy, Zhang, 
\& Kwok 1994).  When the {\it ROSAT} mission concluded in 1998, the 
{\it ROSAT} archive had acquired pointed or serendipitous observations 
with $\ge$2 ks exposures for $\sim$80 PNe.  These observations were 
analyzed, and X-ray emission from 13 PNe was detected (Guerrero, Chu, 
\& Gruendl 2000).  Of these 13 PNe, only three had been reported
to show extended emission that could be associated with hot gas: 
NGC\,6543 (Kreysing et al.\ 1992), A\,30 (Chu \& Ho 1995; Chu, Chang, 
\& Conway 1997), and BD\,+30$^\circ$3639 (Leahy, Kwok, \& Yin 2000). 
In all three cases the X-ray emission was only marginally more 
extended than the point spread function of the detector used.

An indirect indication of diffuse X-ray emission from hot gas in PNe
is provided by the {\it ROSAT} PSPC spectra.  The PNe detected by
$ROSAT$ show three types of spectra (Conway \& Chu 1997; Guerrero, 
Chu, \& Gruendl 2000).  Type 1 spectra, peaking near 0.1--0.2 keV and 
diminishing above 0.5 keV, are consistent with photospheric emission 
from hot CSPNs.  Type 2 spectra, peaking at $\ge$0.5 keV, are 
characteristic for plasma emission.  Type 3 spectra, possessing a 
bright peak near 0.1--0.2 keV and a weak peak at $\sim$1 keV, are 
apparently composites of the first two types.  Objects with Type 2 
spectra are good candidates for diffuse X-ray emission; these include 
BD\,+30$^\circ$3639 and NGC\,6543, and possibly NGC\,7009.  {\it ASCA} 
observations of BD\,+30$^\circ$3639 have provided high-resolution 
spectra and confirmed the nature of plasma emission (Arnaud, 
Borkowski, \& Harrington 1996).

Finally, the superb angular resolution of the {\it Chandra} X-ray 
Observatory made it possible to unambiguously resolve the diffuse 
X-ray emission from PN interiors: BD\,+30$^\circ$3639, NGC\,6543,
and NGC\,7027 (Kastner et al.\ 2000, 2001; Chu et al.\ 2001).
The unprecedented sensitivity of {\it XMM-Newton} made it possible
to resolve the faint diffuse X-ray emission from NGC\,7009
(Guerrero, Gruendl, \& Chu 2002).  These observations allow us
to examine the distribution and physical conditions of the hot gas 
in PN interiors.

Table 2 lists all PNe that have been reported to emit X-rays,
including a PN that was detected after the X-ray source 
(RXJ\,2117.1+3412) and a PN in the SMC.
The possible detections and the mis-identified sources are listed
at the bottom of the table.  The majority of 
the valid X-ray sources in PNe consist of X-ray emission from their
photospheres, showing Type 1 or Type 3 $ROSAT$ PSPC spectra.
Objects showing Type 2 $ROSAT$ PSPC spectra are the best
candidates for diffuse X-ray emission.  References are given for
each object.

{
\begin{table}
\caption{PNe with Reported X-ray Emission}
\vspace*{0.1cm}
\begin{tabular}{lllcl}
\tableline
\tableline
\multicolumn{1}{c}{       } & \multicolumn{1}{c}{            } &
\multicolumn{1}{l}{\it ROSAT}     & 
\multicolumn{1}{c}{       } & \multicolumn{1}{c}{            }  \\ 
\multicolumn{1}{c}{       } & \multicolumn{1}{c}{            } &
\multicolumn{1}{l}{~PSPC}     & 
\multicolumn{1}{c}{Diffuse} & \multicolumn{1}{c}{            }  \\ 
\multicolumn{1}{l}{PN Name} & \multicolumn{1}{l}{Observations} &
\multicolumn{1}{l}{Spectrum} & 
\multicolumn{1}{c}{Emission} & \multicolumn{1}{l}{Reference} \\
\tableline
       &   &   &   &   \\
\multicolumn{5}{c}{\bf PNe with X-ray Sources} \\
       &   &   &   &   \\
 A\,30               & $ROSAT$            & Type 1    & Yes?& 12,17        \\
 A\,36               & $EXOSAT$,$ROSAT$   & Type 2?   &     & 3,19         \\
 BD\,+30$^\circ$3639 & $ROSAT$,$ASCA$     & Type 2    & Yes & 6,14,20,     \\
                     & $Chandra$          &           &     & 21           \\
 K\,1-16             & $ROSAT$            & Type 1    &     & 13,19        \\
 K\,1-27             & $ROSAT$            & Type 2/3? &     & 11           \\
 LoTr\,5             & $EXOSAT$,$ROSAT$   & Type 3    &     & 5,6,13,19    \\
 NGC\,246            & $Einstein$,$EXOSAT$, & Type 1  &     & 2,3,         \\
                     & $ROSAT$            &           &     & 13           \\
 NGC\,1360           & $Einstein$,$EXOSAT$, & Type 1  &     & 1,2,3,       \\
                     & $ROSAT$            &           &     & 13           \\
 NGC\,4361           & $EXOSAT$,$ROSAT$   &           &     & 3,6          \\
 NGC\,3587           & $EXOSAT$,$ROSAT$   & Type 1    &     & 3,15,18      \\
 NGC\,6543           & $ROSAT$,$Chandra$  & Type 2    & Yes & 6,22         \\
 NGC\,6853           & $Einstein$,$EXOSAT$, & Type 1  &     & 2,3,         \\
                     & $ROSAT$            &           &     & 6,8,13       \\
 NGC\,7027           & $Chandra$          &           & Yes & 25           \\
 NGC\,7009           & $ROSAT$,$XMM$      & Type 2    & Yes & 19,26        \\
 NGC\,7293           & $Einstein$,$EXOSAT$,& Type 3   &     & 2,3,         \\
                     & $ROSAT$,$Chandra$  &           &     & 10,15,23,24  \\
 RXJ\,2117.1$+$3412   & $ROSAT$            & Type 1    &     & 7,9          \\
 N67 in the SMC      & $Einstein$         &           &     & 4            \\
       &   &   &   &   \\
\tableline
       &   &   &   &   \\
\multicolumn{5}{c}{\bf PNe with Possible (2$\sigma$) X-ray Sources} \\
       &   &   &   &   \\
 NGC\,2371--72       & $ROSAT$            &           &     & 19           \\
 NGC\,2392           & $ROSAT$            &           &     & 19           \\
 NGC\,6572           & $ROSAT$            &           &     & 19           \\
       &   &   &   &   \\
\tableline
       &   &   &   &   \\
\multicolumn{5}{c}{\bf PNe with Mis-identified X-ray Sources} \\
       &   &   &   &   \\
 A\,12               & $ROSAT$            &           &     & 6,13        \\
 A\,33               & $Einstein$,$ROSAT$ &           &     & 2,16,19     \\
 NGC\,1535           & $EXOSAT$,$ROSAT$   &           &     & 3,15,18     \\
       &   &   &   &   \\
\tableline\tableline
\end{tabular} 
\end{table}
\newpage
{\sc References.---} $\,$ 
 (1) de Korte et al.\ 1985; 
 (2) Tarafdar \& Apparao 1988; 
 (3) Apparao \& Tarafdar 1989;
 (4) Wang 1991; 
 (5) Apparao, Berthiaume, \& Nousek 1992
 (6) Kreysing et al.\ 1992; 
 (7) Appleton, Kawaler, \& Eitter 1993.
 (8) Chu, Kwitter, \& Kaler 1993; 
 (9) Motch, Werner, \& Pakull 1993.
(10) Leahy, Zhang, \& Kwok 1994; 
(11) Rauch, K\"oppen, \& Werner 1994; 
(12) Chu \& Ho 1995;
(13) Hoare et al.\ 1995; 
(14) Arnaud, Borkowski, \& Harrington 1996; 
(15) Leahy et al.\ 1996; 
(16) Conway \& Chu 1997;
(17) Chu, Chang, \& Conway 1997;  
(18) Chu, Gruendl, \& Conway 1998; 
(19) Guerrero, Chu, \& Gruendl 2000; 
(20) Leahy, Kwok, \& Yin 2000; 
(21) Kastner et al.\ 2000; 
(22) Chu et al.\ 2001; 
(23) Gruendl et al.\ 2001;
(24) Guerrero et al.\ 2001; 
(25) Kastner, Vrtilek, \& Soker 2001; 
(26) Guerrero, Gruendl, \& Chu 2002.
}

\subsection{Physical Properties of Hot Gas in PNe}

$Chandra$ and {\it XMM-Newton} observations have resolved diffuse
X-ray emission from four PNe.  The X-ray contour maps and spectra 
of BD\,+30$^\circ$3639, NGC\,6543, and NGC\,7027 are presented in
Figure 1, and those of NGC\,7009 in Figure 1 of Guerrero, Gruendl, 
\& Chu (2002, this volume).   The physical parameters of these
four PNe are summarized in Table 3, where the distance ($d$) and 
size of the nebula, temperature ($T_{\rm e}$) and rms density 
($<$$N_{\rm e}$$>$) of the hot gas, absorption column density 
($N_{\rm H}$), and X-ray luminosity ($L_{\rm X}$) are given.
Despite the uncertainty in distances and the large range of 
absorption column density, it appears that the temperature, 
density, and X-ray luminosity decrease with increasing nebular 
size.  This trend suggests an evolutionary effect: PNe are 
brightest in X-rays when they are young.

\begin{figure}
\plotone{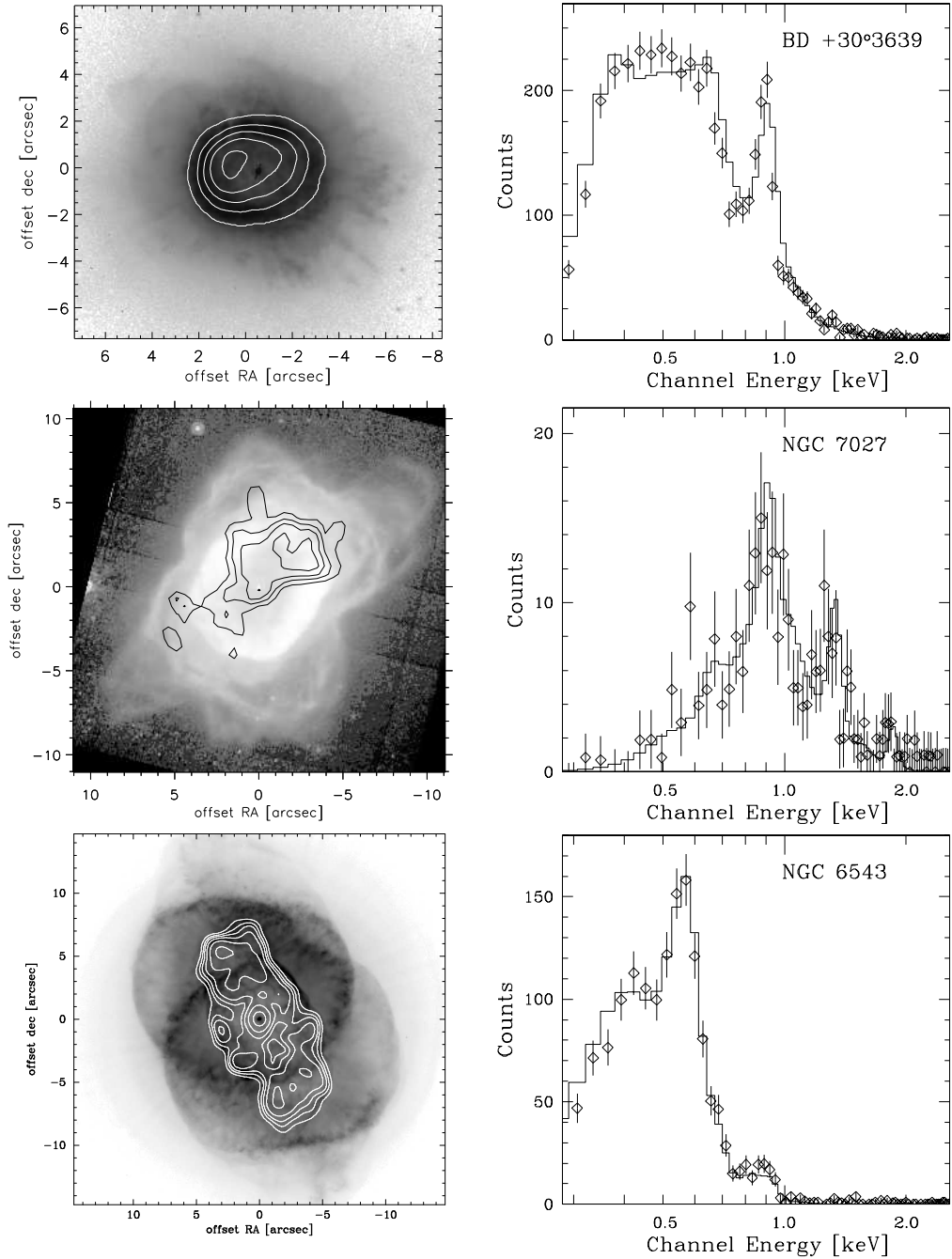}
\caption{$Chandra$ ACIS contour maps and spectra of three PNe,
taken from Kastner et al.\ (2000, 2001) and Chu et al.\ (2001).
The X-ray contours are overlaid on H$\alpha$ images for 
BD\,+30$^\circ$3639 and NGC\,6543, and an H$_2$ 2.12 $\mu$m 
image for NGC\,7027.}
\end{figure}

\vspace*{-0.2cm}
\begin{table}
\caption{Properties of PNe with Extended X-ray Emission}
\vspace*{0.1cm}
\begin{tabular}{lcccccc}
\tableline
\tableline
\multicolumn{1}{c}{PN Name}       & \multicolumn{1}{c}{$d$}         & 
\multicolumn{1}{c}{size}          & \multicolumn{1}{c}{$T_{\rm e}$} & 
\multicolumn{1}{c}{$<N_{\rm e}>$} & \multicolumn{1}{c}{$N_{\rm H}$} & 
\multicolumn{1}{c}{$L_{\rm X}$} \\
\multicolumn{1}{c}{}              & \multicolumn{1}{c}{[kpc]}       &
\multicolumn{1}{c}{[pc]}          & \multicolumn{1}{c}{[$10^6$ K]}  &
\multicolumn{1}{c}{[cm$^{-3}$]}   & \multicolumn{1}{c}{[cm$^{-2}$]} &
\multicolumn{1}{c}{[ergs s$^{-1}$]} \\

\tableline
BD\,+30$^\circ$3639   & 1    & 0.025$\times$0.02   & 2.7   & 200   &
  1$\times$10$^{21}$  & 1.6$\times$10$^{32}$ \\
NGC\,7027             & 0.9  & 0.04$\times$0.03    & 3.0   & 150   &
  6$\times$10$^{21}$  & 1.3$\times$10$^{32}$ \\
NGC\,6543             & 1    & 0.05$\times$0.04    & 1.7   & $\sim$50 &
  9$\times$10$^{20}$  & 1.0$\times$10$^{32}$  \\
NGC\,7009             & 1.2  & 0.145$\times$0.06   & 1.8   & $\sim$25 &
  8$\times$10$^{20}$  & 3$\times$10$^{31}$   \\ 
\tableline
\tableline
\end{tabular} 
\end{table}

\vspace*{-0.3cm}
\section{UV Views of Hot Gas in Planetary Nebulae}

If C$^{+3}$, N$^{+4}$, and O$^{+5}$ are produced by collisional
ionization, the \civ~$\lambda\lambda$1548, 1550, 
\nv~$\lambda\lambda$1238, 1242, and \ovi~$\lambda\lambda$1031, 1037 
lines can be used to diagnose the presence of $1\times10^5$ K, 
$2\times10^5$ K, and $3\times10^5$ K gas, respectively.  
Note, however, that C$^{+2}$, N$^{+3}$, and O$^{+4}$, having
ionization potentials of 47.9 eV, 77.5 eV, and 113.9 eV,  
can be photoionized by stars with effective temperatures higher 
than 35,000~K, 75,000~K, and 125,000~K, respectively.  Therefore, 
UV absorption line studies of 10$^5$ K gas need to be carefully 
designed to minimize the confusion caused by photoionization.

High-dispersion UV observations of nebular \civ\ and \nv\ absorption 
lines in PNe have been made with the {\it International Ultraviolet 
Explorer (IUE)} and {\it Hubble Space Telescope (HST)}.  For example, 
$IUE$ observations of NGC\,6543 (Pwa, Mo, \& Pottasch 1984) and $HST$
GHRS observations of A\,78 (Harrington, Borkowski, \& Tsvetanov 1995)
showed nebular \civ\ absorption.  As the He~{\sc ii} or H Zanstra 
temperatures of their central stars are $\sim$50,000 K and $\sim$70,000
K, respectively (Kaler \& Jacoby 1991; Kaler 1983), these 
observations are not useful for the study of 10$^5$ K gas. \nv\ lines
have small oscillator strengths, so are difficult to detect.

\ovi\ lines in the far-UV, being strong and less affected by 
photoionization, are the best lines to study 10$^5$ K gas. 
High-dispersion spectroscopic observations of the \ovi\ lines
have been made with the {\it Orbiting Retrievable Far and Extreme 
UV Spectrometer (ORFEUS)} for only the brightest CSPNs, e.g.,
NGC\,6543 (Zweigle et al.\ 1997).  The spectral resolution of 
$ORFEUS$, $\lambda/\Delta \lambda$ = 3000, was not adequate to 
distinguish clearly among the interstellar and nebular \ovi\ and 
H$_2$ absorption lines.  It was not clear whether nebular \ovi\ 
absorption was present in the spectrum of NGC\,6543.

Finally, the {\it Far-UV Spectroscopic Explorer (FUSE)} (Sonneborn 2002)
provides the sensitivity and resolution required to study the nebular
\ovi\ absorption and emission lines for a large number of PNe.  We have
initiated $FUSE$ programs to study the hot gas at the interfaces in
PNe.  We have obtained $FUSE$ observations for seven CSPNs for a wide
range of stellar temperatures.  The spectral region around the \ovi\ 
lines of these observations is presented in Figure 2.  This figure
demonstrates clearly that the stellar and interstellar/nebular lines 
are very complex.  The spectra of NGC\,2392 and NGC\,6058 show narrow 
absorption components at expected locations for both 
\ovi~$\lambda\lambda$1031, 1037 lines, and thus provide the most
promising candidates for detections of the 10$^5$ K gas at the
interface.  We are in the process of enlarging our sample to include 
all archival $FUSE$ observations of CSPNs in order to empirically sort 
out the temperature dependence of the spectral features of the stars
and to pinpoint the nebular absorption components (Gruendl et al.,
in preparation).  

\begin{figure}
\plotone{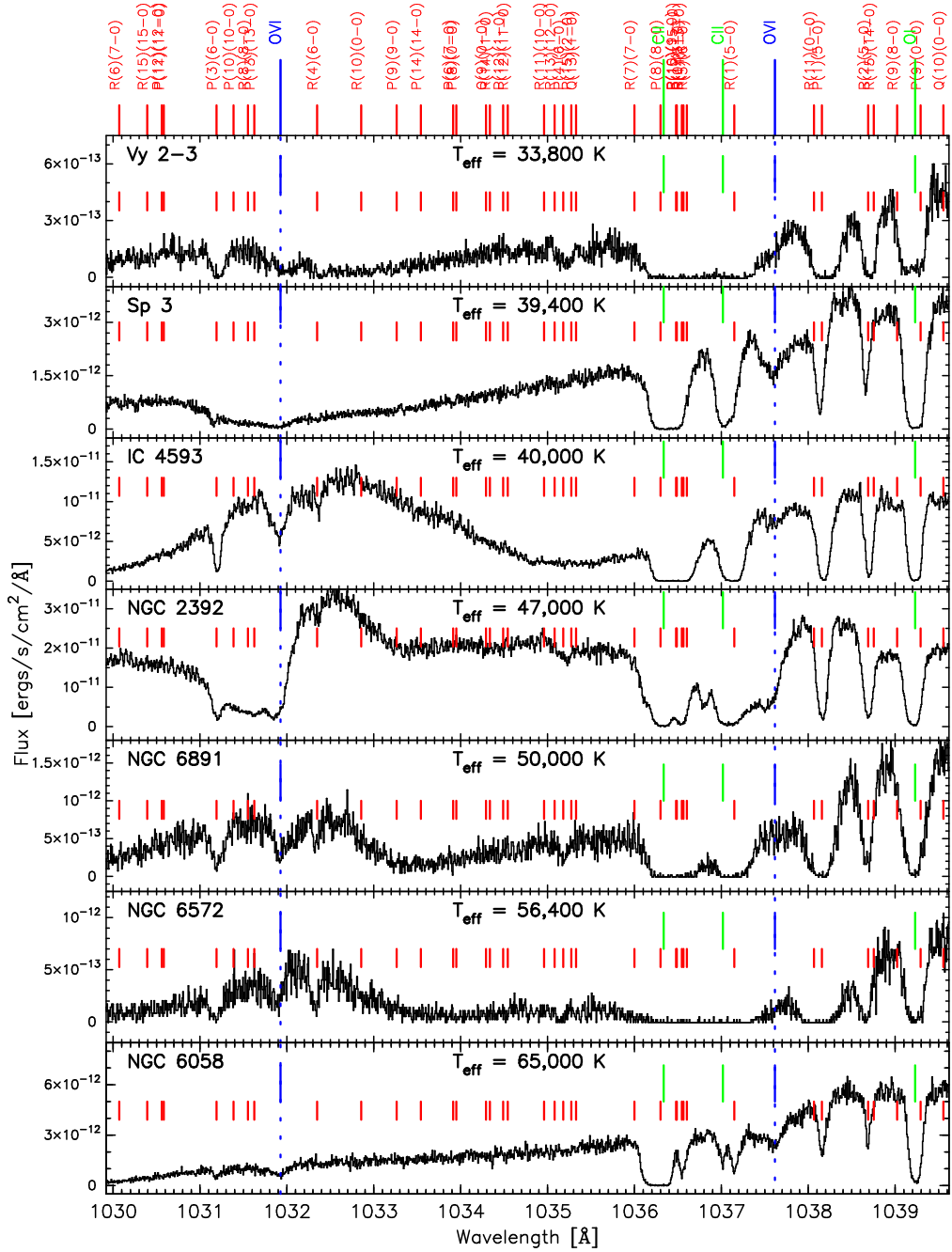}
\caption{$FUSE$ spectra of seven CSPNs in the \ovi\ lines.  The 
spectra are arranged according to the stellar $T_{\rm eff}$.  The
location of \ovi, C~{\sc ii}, O~{\sc i}, and H$_2$ lines are marked
on the top margin.}
\end{figure}

\section{Concluding Remarks}

$Chandra$ and {\it XMM-Newton} observations have unambiguously detected
diffuse X-ray emission from the hot gas in PN interiors, illustrating 
that PNe are excellent laboratories to study shocked stellar winds
and the interfaces between this hot gas and cool nebular shells.
Preliminary results show that the hot gas is overpressurized
and drives the nebular expansion.  
It is puzzling, however, that the shocked stellar winds in PNe can  
cool from 10$^7$--10$^8$ K to 2--3$\times$10$^6$ K within the short 
dynamic age of the nebular shells, $\sim$1,000 yrs.
Future $Chandra$ and {\it XMM-Newton} observations of hot interior gas,
in conjunction with $FUSE$ observations of the interfaces, for PNe with
a wide range of morphologies and evolutionary stages are needed to
empirically establish our knowledge of the physical conditions and 
evolution of the hot gas in PN interiors.  
A sound knowledge of the hot gas in PN interiors is crucial to
our understanding of the formation and evolution of PNe as a 
whole.

\acknowledgements
We gratefully acknowledge the observing times we received from $Chandra$,
{\it XMM-Newton}, and {\it FUSE}.  This work has been supported by NASA
grants associated with these observing programs: SAO~GO~0-1004X,
NAG 5-10042, and NAG 5-10182.


\newpage
\begin{references}
\reference \arg{Apparao, K.\ M.\ V., \& Tarafdar, S.\ P.\ 1989, \apj, 344, 826}
\reference \arg{Apparao, K.\ M.\ V., Berthiaume, G.\ D., \& Nousek, J.\ A.\
    1992, \apj, 397, 534}
\reference \arg{Appleton, P.\ N., Kawaler, S.\ D., \& Eitter, J.\ J.\ 1993,
    \aj, 106, 1973}
\reference \arg{Arnaud, K., Borkowski, K.\ J., \& Harrington, J.\ P.\ 1996, 
                \apj, 462, L75}
\reference \arg{Chu, Y.-H., Chang, T.\ H., \& Conway, G.\ M.\ 1997, \apj, 482, 
                891}
\reference \arg{Chu, Y.-H., Gruendl, R.\ A., \& Conway, G.\ M.\ 1998, \aj, 116, 
                1882}
\reference \arg{Chu, Y.-H., Guerrero, M.\ A., Gruendl, R.\ A., Williams, R.\ 
                M., \& Kaler, J.\ B.\ 2001, \apjl, 553, L69}
\reference \arg{Chu, Y.-H., \& Ho, C.-H. 1995, \apj, 448, L127}
\reference \arg{Chu, Y.-H., Kwitter, K., \& Kaler, J.\ 1993, \aj, 106, 650}
\reference \arg{Conway, G.\ M., \& Chu, Y.-H.\  1997, in IAU Symp.\ 180, 
                Planetary Nebulae, ed.\ H.\ J.\ Habing, \& H.\ J.\ G.\ L.\ M.\
                Lamers (Dordrecht: Kluwer), 214}
\reference \arg{de Korte, P.\ A.\ J., Claas, J.\ J., Jansen, F.\ A., \& 
                McKechnie, S.\ P.\ 1985, Adv.\ Space Res., 5, 57}
\reference \arg{Gruendl, R.\,A., Chu, Y.-H., O'Dwyer, I.\,J., \& Guerrero, M.\ 
     2001, \aj, 122, 308}
\reference \arg{Guerrero, M.\ A., Chu, Y.-H., \& Gruendl, R.\ A.\ 2000, \apjs, 
                129, 295}
\reference \arg{Guerrero, M.\ A., Chu, Y.-H., Gruendl, R.\ A., Williams, R.\ 
                M., \& Kaler, J.\ B.\ 2001, \apjl, 553, L55}
\reference \arg{Guerrero, M.\ A., Gruendl, R.\ A., \& Chu, Y.-H.\ 2002, 
      \aap, in press; also in this volume}
\reference \arg{Harrington, J.\ P., Borkowski, K.\ J., \& Tsvetanov, Z.\
     1995, \apj, 439, 264}
\reference \arg{Hoare,\,M.G., Martin,\,A.B.,\, Werner,\,K., \&
  Fleming,\,T.\ 1995, \mnras, 273, 812}
\reference \arg{Kaler, J.\ B.\ 1983, \apj, 271, 188}
\reference \arg{Kaler, J.\ B.,\ Jacoby, G.\ H.\ 1991, \apj, 372, 215}
\reference \arg{Kastner, J.\ H., Soker, N., Vrtilek, S.\ D., \& Dgani, R.\
       2000, \apjl, 545, L57} 
\reference \arg{Kastner, J.\ H., Vrtilek, S.\ D., \& Soker, N.\ 2001, \apjl, 
               550, L189} 
\reference \arg{Kreysing, H.\ C., Diesch, C., Zweigle, J., Staubert, R., 
                Grewing, M., \& Hasinger, G.\ 1992, \aap, 264, 623}
\reference \arg{Kwok, S., Purton, C.\ R., \& FitzGerald, P.\ M.\ 1978,
  \apj, 219, L125}
\reference \arg{Leahy, D.\ A., Kwok, S., \& Yin, D.\ 2000, \apj, 540, 442}
\reference \arg{Leahy, D.\ A., Zhang, C.\ Y., \& Kwok, S.\ 1994, \apj, 422, 
                L205}
\reference \arg{Leahy, D.\ A., Zhang, C.\ Y., Volk, K., \& Kwok, S.\ 1996, 
                \apj, 466, L352}
\reference \arg{Motch, C., Werner, K., \& Pakull, M.\ W.\ 1993, \aap, 268, 561}
\reference \arg{Pwa, T.\ H., Mo, J.\ E., \& Pottasch, S.\ R.\ 1984,
      \aap, 139, L1}
\reference \arg{Rauch, T., K\"oppen, J., \& Werner, K.\ 1994, \aap, 286, 543}
\reference \arg{Sonneborn, G.\ 2002, in this volume}
\reference \arg{Tarafdar, S.\ P., \& Apparao, K.\ M.\ V.\ 1988, \apj, 327, 342}
\reference \arg{Wang, Q.\ 1991, \mnras, 252, 47p}
\reference \arg{Weaver,\,R., McCray,\,R., Castor,\,J., Shapiro,\,P., Moore,\,R.\
    1977, \apj, 218, 377}
\reference \arg{Zweigle, J.\ et al.\ 1997, \aap, 321, 891}
\end{references}
\end{document}